\documentclass{JHEP}
\usepackage{epsfig}
\usepackage{amssymb}
\usepackage{amsmath}
\newcommand{\be}{\begin{equation}}
\newcommand{\ee}{\end{equation}}
\def\bea{\begin{eqnarray}}
\def\eea{\end{eqnarray}}

\newcommand{\A}[1]{A^{(#1)}}

\newcommand{\G}[1]{G^{(#1)}}
\newcommand{\CHI}[1]{\chi^{(#1)}}

\newcommand{\PSI}[1]{\Psi^{(#1)}}

\newcommand{\wdg}{{\scriptscriptstyle \wedge}}

\title{Domain Walls
with Strings Attached}

\author{Renata Kallosh\\
Department of Physics, Stanford University, Stanford, CA 94305, USA\\
    E-mail: \email{kallosh@stanford.edu}}
\author{Sergey Prokushkin\\
    Department of Physics, Stanford University, Stanford, CA 94305,
USA\\
    E-mail: \email{prok@stanford.edu}}
    \author{Marina Shmakova\\
    CIPA, 366 Cambridge Avenue, Palo Alto, CA 94306 and \\
   SLAC, Stanford University, Stanford, CA 94309 \\  \email{shmakova@slac.stanford.edu}}
\received{\today}       
 \preprint{SU-ITP-01-33\\ SLAC-PUB-8915\\ \hepth{0107097}\\ \today}

\abstract{We have constructed a bulk \& brane
action of IIA theory which describes a pair of BPS domain walls on
${S_1/ \bf {Z}_2}$, with strings attached. The walls are given
by  two orientifold O8-planes with coincident D8-branes and `F1-D0'-strings
are stretched between the walls. This static configuration  satisfies all
matching conditions for the string and domain wall sources and has 1/4
of unbroken supersymmetry.}

\keywords{eld.pbr.ctg.sgm}

\begin{document}


Recently a new approach to type IIA theory in 10d was initiated in \cite{BerKalOrDP:2001}. It was given a name `bulk \& brane action' \footnote{This was a follow up of a 5d bulk \& brane action suggested in \cite{Bergshoeff:2000zn} with the purpose to supersymmetrize the Randall-Sundrum \cite{RS} brane world. This approach to IIA theory is closely related to
 Polchinski-Witten construction \cite{Polchinski:1996df}.}. The purpose was to clarify the properties of D8 branes (domain walls in 10d)  and related configurations which require the `massive' type IIA bulk supergravity of Romans \cite{Romans:1986}.
The new version of IIA supergravity in the bulk \cite{BerKalOrDP:2001} includes a 10-form field strength $G^{(10)}$ which is dual to zero-form field strength $G^{(0)}$. The theory  is defined on ${S_1/ \bf {Z}_2}$ space. The value of $G^{(0)}$ depends on the brane part of the bulk \& brane action. The brane part consists of two orientifold planes and some number of D8 branes coincident with each of the O8 planes. The resulting {\it on shell value of the 0-form field $G^{(0)}$  in the presence of sources  is a piecewise constant}
$$
G^{(0)}(x^9)= {2n-16\over 4\pi l_s}\epsilon (x^9)
\label{G0}\, .
$$
This identifies the mass parameter of Type~IIA supergravity in `bulk \& brane action' \cite{BerKalOrDP:2001} as
follows:
\begin{equation} 
m= \left\{
  \begin{array}{cc}
 {\displaystyle\frac{n-8}{2\pi\ell_{s}}}\, ,\,\,\,\, & 0< x^9 < \pi R\, ,\\
& \\
- {\displaystyle\frac{n-8}{2\pi\ell_{s}}}\, ,\,\,\,\, & -\pi R< x^9 < 0\, .\\
  \end{array}
\right. \label{quant}\nonumber
\end{equation}
The mass is quantized in string units and it is proportional to $n-8$
where there are $2n$ and $2(16-n)$ D$8$-branes (including the images) at each O$8$-plane. The
mass vanishes  in the special case $n=8$ when the contribution from
the D$8$-branes cancels exactly the contribution from the O$8$-planes.
$$
n=8 \qquad  \Rightarrow \qquad m=0 \ .
\label{m=0}
$$
The set up for the bulk \& brane action allows to find the equations of motion following from the bulk supergravity as well as from the brane actions. A familiar example of this kind is  the fundamental string F1: the equations of motion were derived  from the bulk 10d massless supergravity supplemented by the string action \cite{Dabholkar:1990yf}. The coordinates of the string were embedded into the target space of supergravity by the choice of the static gauge: $X^0=\tau, X^1=\sigma$.
The matching conditions at the position of the string were satisfied for this solution. In \cite{BerKalOrDP:2001} an analogous procedure was performed for stringy domain walls in massive IIA bulk supergravity supplemented by O8 and D8 actions. The solution satisfied matching conditions at the positions of the walls and had 1/2 of unbroken supersymmetry.

The purpose of this paper is to introduce more general `bulk \& brane actions' and study the solutions of massive IIA theory with less unbroken supersymmetry. The challenge here is that apart from D8 branes
which require massive supergravity, the rest of the known D-branes are solutions of the massless supergravity. Few solutions of the massive supergravity have been found before \cite{d0d8f1Ort,d0d8f1}. However they have been found in the theory with the constant mass, without a 10-form,  whereas a consistent theory of domain walls on ${S_1/ \bf {Z}_2}$ requires the presence of the 10-form dual to a 0-form and they both  change the sign across the wall.

An interesting case which we will study in this paper is  a configuration which one can conditionally call D8-D0-F1 solution which is expected to define strings stretched between the walls. One can not expect a simple combination of known D8 brane, D0 brane and fundamental strings F1 solutions by the following reason:

\begin{itemize}

\item The status of the D0 brane in massive supergravity with  everywhere constant mass is somewhat ambiguous
since the 1-form can be gauged away via Higgs effect which makes the 2-form $B$ a massive field \cite{Romans:1986}. In our theory \cite{BerKalOrDP:2001}, however, in presence of the 0- and 10-form fields it is impossible to gauge the RR 1-form away everywhere.
In  Romans theory \cite{Romans:1986} the field $B_{\mu\nu}$ appears either via the field strength $H=dB$ or in the combination $dA + mB$. Thus the transformation  $\delta B= {1\over m} d A$ does not change $H$ and absorbs the 1-form $A$ into $B$.  In  \cite{BerKalOrDP:2001} $B$ enters in a combination  $dA +  G^{(0)}B$ where the 0-form $G^{(0)}$ is a function and therefore $G^{(0)}B$ can not absorb $dA$, in general.

\item The fundamental string F1 of type IIA massless supergravity does not solve equations of motion of the massive theory. Therefore in presence of O8-D8 walls some unusual strings may be expected.

\end{itemize}

Our interest to the problem was enhanced also by a phenomenon of {\it string creation when D0 particle crosses a D8 brane} \cite{creation}. The observation in these papers was of the following nature. One considers one isolated D8-brane and one assumes that on one side of the 
D8-brane, for example the right hand side,  there is a bulk defined  by  massless supergravity, 
$$
m_{\rm rhs}=0 \qquad \Rightarrow \qquad D0 \ .
\label{m=0}
$$
Therefore on the right hand  side of the D8-brane the usual D0-branes are possible since they solve equations of motion of massless supergravity. On the other side of the D8-brane the bulk is assumed to have a non-vanishing mass, i. e. the bulk is defined by the massive supergravity. The usual D0-brane can not exist there without a $B_{\mu\nu}$-field. Thus a string must be created as soon as the D0-brane crosses the D8-branes and appears on the other side where 
$$
m_{\rm lhs} \neq 0 \qquad \Rightarrow \qquad D0 + F1 \ .
$$
In our case when we have two O8-D8 domain walls at the fixed points of the orientifold, the mass changes the sign across the wall, but is nowhere vanishing. Therefore we expect to find a solution which everywhere has a modified string combined with some modified charged D-particle.

We will  start  from d=10 Lagrangian for dual IIA which is given in eq. (2.23) of  \cite{BerKalOrDP:2001}
 with  the independent fields \footnote{We keep all notation of \cite{BerKalOrDP:2001}.}.
\begin{equation}
  \left\{   e_\mu ^a,
  B_{\mu \nu},
  \phi,
  \G{0},
  \G{2}_{\mu \nu },
  \G{4}_{\mu _1\cdots \mu _4},
  \A{5}_{\mu _1\cdots\mu _5 },
  \A{7}_{\mu _1\cdots\mu _7 },
  \A{9}_{\mu _1\cdots\mu _9 },
  \psi_\mu,
  \lambda
  \right\}.
 \label{IIAFields}
\end{equation}
The bulk action is
\bea
 S_{\rm{bulk}}^{5,7,9} =
  & - \frac{1}{2\kappa_{10}^2}\int d^{10} x \sqrt{-g}
    \Big\{
    e^{-2\phi} \big[
    R\big(\omega(e)\big) -4\big( \partial{\phi} \big)^{2}
    +\frac{1}{2} H \cdot H
    -2\partial^{{\mu}}{\phi} \CHI{1}_{{\mu}}
        + H \cdot \CHI{3}+ \nonumber \\
  & +2 \bar{{\psi}}_{{\mu}}{\Gamma}^{{\mu}{\nu}{\rho}}
    {\nabla}_{{\nu}}{\psi}_{{\rho}}
    -2 \bar{{\lambda}}{\Gamma}^{{\mu}}
    {\nabla}_{{\mu}}{\lambda}
    +4 \bar{{\lambda}} {\Gamma}^{{\mu}{\nu}}
    {\nabla}_{{\mu}}{\psi}_{{\nu}}
    \big]
    +   \sum_{n=0,1,2}
     \frac{1}{2} \G{2n} \cdot \G{2n}
    + \G{2n} \cdot \PSI{2n}  +  \nonumber \\
  & - \star \, \big[
      \frac{1}{2} \, \G{4} \G{4} B
    - \frac{1}{2} \, \G{2} \G{4} B^2
    + \frac{1}{6} \, \G{2}{}^2 B^3
    + \frac{1}{6} \, \G{0} \G{4} B^3 -\frac{1}{8} \,
    \G{0} \G{2} B^4+ \nonumber \\
  & +\frac{1}{40} \, \G{0}{}^2 B^5
    + {\bf e}^{- B} {\bf G}
    d (\A{5} - \A{7} + \A{9}) \big] \Big\}
    +\mbox{ quartic fermionic terms}\,,
    \label{dualaction}
\eea
where  ${\bf G} =\sum_{n=0}^5 G^{(2n)}$ is a formal sum and
\bea
  \CHI{1}_\mu &=&
  -2 \bar{\psi}_\nu \Gamma^\nu \psi_\mu
    -2 \bar{\lambda} \Gamma^\nu \Gamma_\mu \psi_\nu\, , \nonumber\\
  \CHI{3}_{\mu\nu\rho} &=
  & {\textstyle \frac{1}{2} }\bar{{\psi}}_{{\alpha}}
    {\Gamma}^{[{\alpha}}
    \Gamma_{\mu\nu\rho}
    {\Gamma}^{{\beta}]}
    {\cal P}{\psi}_{{\beta}}
    + \bar{{\lambda}}
    \Gamma_{\mu\nu\rho}{}^{\beta}
    {{\cal P}}{\psi}_{{\beta}}
    -{\textstyle\frac{1}{2}}\bar{{\lambda}}
     {\cal P}\Gamma_{\mu\nu\rho}
    {\lambda}\, , \nonumber \\
  \PSI{2n}_{\mu_1\cdots \mu_{2n}} &=
  & {\textstyle\frac{1}{2}}e^{-{\phi}}
    \bar{{\psi}}_{{\alpha}}
    {\Gamma}^{[{\alpha}}
    \Gamma_{\mu_1\cdots \mu_{2n}}
    {\Gamma}^{{\beta}]}
    {\cal P}_n {\psi}_{{\beta}}
    +{\textstyle\frac{1}{2}}e^{-{\phi}}
    \bar{{\lambda}}
    \Gamma_{\mu_1\cdots \mu_{2n}}
    {\Gamma}^{{\beta}}
    {\cal P}_n{\psi}_{{\beta}}+ \nonumber\\
  & - &{\textstyle\frac{1}{4}}
    e^{-{\phi}}
    \bar{{\lambda}}
    \Gamma_{ [ \mu_1\cdots \mu_{2n-1}}
    {\cal P}_n \Gamma_{\mu_{2n} ] } {\lambda}\, .\nonumber
\eea
The fields $\G{0},
  \G{2}_{\mu \nu },
  \G{4}_{\mu _1\cdots \mu _4}$ are auxiliary since they enter into the action without derivatives and can be integrated out so that the action will depend on the field strength of  $\A{5}_{\mu _1\cdots\mu _5 },
  \A{7}_{\mu _1\cdots\mu _7 },
  \A{9}_{\mu _1\cdots\mu _9 }$ R-R forms. Alternatively, it was explained   in \cite{BerKalOrDP:2001}, one can dualize this action and bring it to the form closely related to the original action of Romans \cite{Romans:1986} where the 1-form $C^{(1)}_\mu$ and the 3-form $C^{(3)}_{\mu\nu\lambda}$ appear and the mass is constant. The change of the basis for RR-forms must be performed for such transition,   $ \mathbf{A}= \mathbf{C}\wedge e^{-B}$.  The action in stringy frame is given in eq. (2.33) of \cite{BerKalOrDP:2001}.

None of these actions, neither the  one in (\ref{dualaction}) nor the Romans-type action, can be used directly to find a solution we are looking for, since we need both  a 1-form for D0 and a 9-form for O8-D8 wall. However, it is easy to bring the action (\ref{dualaction}) to a desirable form.

Our goal therefore  is to construct a partially dual  Lagrangian in terms of
 independent fields
\begin{equation}
  \left\{   e_\mu ^a,
  B_{\mu \nu},
  \phi,
  \G{0},\G{4}_{\mu _1\cdots \mu _4},
  \A{1}_{\mu},
  \A{5}_{\mu _1\cdots\mu _5 },
  \A{9}_{\mu _1\cdots\mu _9 },
  \psi_\mu,
  \lambda
  \right\}
 \label{048Fields}
\end{equation}
that will correspond to the D0-F1-D8 brane system.
We can
express the auxiliary field $\G{2}$ via $\A{1}$, $\G{0}$ and $B$ using the field equations for  $\A{7}$ following from (\ref{dualaction})
$$
-\int {\bf e}^{- B} {\bf G}\wdg d\A{7} = \int
  (d{\bf e}^{- B} {\bf G})\wdg \A{7} + d[\cdots] \qquad \Rightarrow \qquad d(\G{2} - \G{0} B)=0
$$
The most general solution is \cite{BerKalOrDP:2001}:
 $\G{2} = d \A{1} + \G{0} B + \G{2}_{\rm flux},$ where $d\G{2}_{\rm flux}=0$.
 We will choose $\G{2}_{\rm flux}=0$ and will substitute the solution
\bea
{\Large  \G{2}(A^1, \G{0}, B) = d \A{1} + \G{0} B} \label{G2}
\eea
into (\ref{dualaction}) which will give us a partially dual action we are looking for :
\bea
\label{partialdual}
 S_{\rm{bulk}}^{1,5,9}& = &
   - \frac{1}{2\kappa_{10}^2}\int d^{10} x \sqrt{-g}
    \Big\{
    e^{-2\phi} \big[
    R\big(\omega(e)\big) -4\big( \partial{\phi} \big)^{2}
    +\frac{1}{2} H \cdot H
    -2\partial^{{\mu}}{\phi} \CHI{1}_{{\mu}}
        + H \cdot \CHI{3}+ \nonumber \\
  & + &2 \bar{{\psi}}_{{\mu}}{\Gamma}^{{\mu}{\nu}{\rho}}
    {\nabla}_{{\nu}}{\psi}_{{\rho}}
    -2 \bar{{\lambda}}{\Gamma}^{{\mu}}
    {\nabla}_{{\mu}}{\lambda}
    +4 \bar{{\lambda}} {\Gamma}^{{\mu}{\nu}}
    {\nabla}_{{\mu}}{\psi}_{{\nu}}
    \big] \nonumber \\
  &+ & \frac{1}{2} \G{0} \cdot \G{0}
    + \frac{1}{2} \G{4} \cdot \G{4}
    + \frac{1}{2}(d \A{1}+\G{0} B)(d \A{1}+\G{0} B)  
   \label{dualaction1} \\
  & +& \G{0}\PSI{0}
    + (d \A{1}+\G{0} B)\cdot \PSI{2} 
    + \G{4}\cdot \PSI{4}   \nonumber \\
  & -& \star \, \big[
      \frac{1}{2} \, \G{4} \G{4} B
    - \frac{1}{3} \, \G{0} \G{4} B^3
    + \frac{1}{15} \, \G{0}{}^2 B^5
    + \frac{1}{6}d \A{1}d \A{1} B^3 \nonumber \\
  &+ &(\frac{5}{24} \G{0} B^4
    - \frac{1}{2}\G{4} B^2)d \A{1}
   +(\G{4} -  B d \A{1} 
   - \frac{1}{2} \G{0} B\wdg B)d \A{5} 
    +\G{0}d \A{9} 
  \big] \Big\}  \nonumber \\
  & + &\mbox{ quartic fermionic terms}\,. \nonumber
 \eea
Substitution of (\ref{G2}) into supersymmetry transformation rules found in \cite{BerKalOrDP:2001} for the action (\ref{dualaction}) gives the supersymmetry transformations of our new partially dual action  (\ref{partialdual}):
\bea
\delta_{{\epsilon}} {\psi}_{{\mu}} & =
  & \Big( \partial_{{\mu}} +{\textstyle\frac{1}{4}}
    \not\!{\omega}_{{\mu}}
    +{\textstyle\frac{1}{8}}\Gamma_{11}\not\!\! {H}_{\mu}
    \Big) {\epsilon}
    +{\textstyle\frac{1}{8}} e^{{\phi}} \Big({G}^{(0)} {\Gamma}_{{\mu}}+
    {\textstyle\frac{1}{2}}(2 \, \partial_{[\nu}\A{1}_{\rho]}
    {\Gamma}^{\nu \rho}
     +\G{0}\not\!B){\Gamma}_{{\mu}} {\Gamma}_{11}
    +{\textstyle\frac{1}{24}} \not \! {G}^{(4)} {\Gamma}_{{\mu}}
    \Big){\epsilon}\, , \nonumber \\
     \delta_{{\epsilon}}{\lambda} & =
    & \Big( \! \! \not \! \partial \phi
    - {\textstyle\frac{1}{12}} {\Gamma}_{11} \not\!\! {H} \Big)
      {\epsilon}
    + {\textstyle\frac{1}{4}} e^{{\phi}} \Big(5 {G}^{(0)}
    + {\textstyle\frac{3}{2}}
    (2 \, \partial_{[\nu}\A{1}_{\rho]}{\Gamma}^{\nu \rho}
    +\G{0}\not\!B){\Gamma}_{11}
     +{\textstyle\frac{1}{24}} \not \! {G}^{(4)}
     \Big){\epsilon} \,, \nonumber\\
     \delta_{{\epsilon}}{\phi} &= &
      {\textstyle\frac{1}{2}} \, \bar{{\epsilon}}{\lambda} \,, \nonumber\\
     \delta\G{0} &=& 0\, , \nonumber \\
       \delta \A{1}&=& -e^{-{\phi}}\bar{\epsilon}{\Gamma}_{11}
     \left( {\psi}_{{\mu}}
    -{\textstyle\frac{1}{2}}{\Gamma}_{{\mu}}{\lambda}\right)\, , \nonumber \\
    \delta\G{2}&=&dE^1
    +\G{0}\wdg \delta_{\epsilon}B = \delta (d \A{1}+\G{0}B)\, , \nonumber \\
     \delta\G{4}&=&dE^3
    +\G{2}\wdg\delta_{\epsilon}B-H\wdg E^1 \, , \nonumber \\
    \delta \A{5}&=&E^5
    -B\wdg E^3+{\textstyle\frac{1}{2}}B\wdg B \wdg E^1 \, ,  \nonumber \\
      \delta \A{9}&=&E^9-B\wdg E^7
     +{\textstyle\frac{1}{2}}B\wdg B \wdg E^5
    -{\textstyle\frac{1}{6}}B \wdg B \wdg B \wdg E^3 +
     {\textstyle\frac{1}{24}}B \wdg B \wdg B \wdg B \wdg E^1 \, ,
\eea
where
$$ E^{(2n-1)}_{\mu_1 \cdots \mu_{2n-1}} \equiv
    - e^{-\phi} \, \bar{\epsilon} \,
    \Gamma_{[\mu_1\cdots \mu_{2n-2}} \, (\Gamma_{11})^n \,
   \Big((2n-1)\psi_{\mu_{2n-1}]}
  - {\textstyle\frac{1}{2}} \Gamma_{\mu_{2n-1}]}
    \lambda\Big)\,. $$
We now make an assumption that our solution has 
$\G{4}=0$, $\A{5}=0$, $B\wedge B=0$, $dA\wedge B=0$, $dA \wedge dA=0$.
The full action whose variation will define the D0-D8-F1 solution will consist of the bulk action and source actions. The brane source action for domain walls was presented in \cite{BerKalOrDP:2001}. The sources for F1-D0 are not known and we hope to find them when the bulk solution will be established. Thus we take
\begin{equation}
 S_{\rm bulk \& brane} = S_{\rm bulk} + S_{O8D8} + S_{F1D0}\, .
\label{totalaction}
\end{equation}
The simplified form of the bulk bosonic action (\ref{partialdual}) which we need for our solution is: 
\bea
  \label{bulk}
 S_{\rm{bulk}}& = &
   - \frac{1}{2\kappa_{10}^2}\int d^{10} x \sqrt{-g}
    \Big\{
    e^{-2\phi} \big[
    R\big(\omega(e)\big)
 -4\big( \partial{\phi} \big)^{2}
    + \frac{1}{12}( H_{\mu\nu\lambda})^2 \big]  \nonumber \\
 &+ & \frac{1}{2} (\G{0})^2 
    + \frac{1}{2}(d \A{1}+\G{0} B)^2- \star \, \big[
    \G{0}d \A{9}
  \big] \Big\} \, .
 \eea
The action of O8D8 (which is a D8-brane action in a static gauge, with excitations on the brane frozen)
is 
\begin{eqnarray}
S_{O8+D8}&=&-
\mu_8
{\displaystyle\int} d^{10}{x}\,  \{  e^{-\phi}
\sqrt{|g_{(9)}|}
 +\alpha \frac{1}{9!} \varepsilon^{(9)} \A{9} \}
  \left(\delta(z)- \delta(z- \pi R)\right)\, , \label{8O+8D}
\label{O+D}
\end{eqnarray}
where $\mu_8= \frac{1}{2\kappa_{10}^2}\frac{2 (n-8)}{(2\pi \ell_s)}$. Here $-16$ is the contribution to the tension from each orientifold plane. The positive part of the tension comes from  D8 branes coincident with the O-planes.

\begin{figure}[htbp]
\centering
\epsfig{file=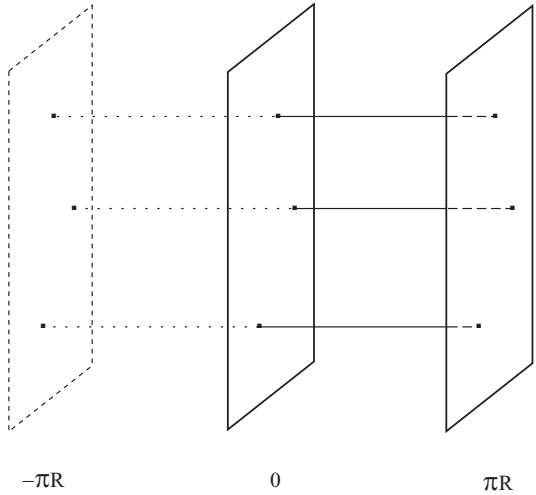, width=7cm}
\caption{At $z=0$ there is an O8-plane with $2n$ D8-branes on it, at $|z|=\pi R$ there is a second O8-plane with $2(16-n)$ D8-branes on it (counting the images). The configuration includes an orthogonal to domain walls `F1-D0' collection of strings sitting at positions $\vec x_k$, all in static equilibrium, 1/4 of supersymmetry unbroken. }
\label{fig:example}
\end{figure}

We will find the following   D0-D8-F1 solution\footnote{Our ansatz is motivated by the results from \cite{d0d8f1} where a closely related solution was obtained in Romans theory \cite{Romans:1986} with everywhere constant mass. On the other hand, for $\G{0}(x)= -\star  \G{10}(x)$,  when the string harmonic function $f$ is trivial,  $N_k=0$ and $ f=1$, our solution is reduced to the one for two O8-D8 domain walls on $S^{1}/\mathbb{Z}_{2}$ derived in \cite{BerKalOrDP:2001}.
} on $S^{1}/\mathbb{Z}_{2}$.
It  depends on 2 harmonic functions, $h(z)$, where 
$z$ is a coordinate transverse to the D8 brane,
and $f(x^i)$ where $x^i$ are the 8 coordinates transverse to the string.
\bea
  ds^2 &=& -h^{-\frac{1}{2}}f^{-\frac{3}{2}}dt^2
  + h^{\frac{1}{2}}f^{-\frac{1}{2}}dz^2
  + h^{-\frac{1}{2}}f^{\frac{1}{2}}
  (dx^i)^2 \label{metr1}  \\
  \nonumber\\
  e^{\phi} &=&  h^{-\frac{5}{4}}f^{\frac{1}{4}} \, \qquad 
  B_{zt} = a f^{-1} \ , \qquad 
  A_t = b hf^{-1}     \label{a1}  \\
   \nonumber\\
  \G{2}_{it}& = & b h \partial_i f^{-1}\ ,  \qquad  \G{2}_{zt}=0 
  \\
    \nonumber\\
 \G{0}&=& -\star  \G{10}= \alpha {n-8\over 2\pi l_s} \epsilon (z)= - \alpha \partial_z h 
 \label{g21}\eea
 where $\alpha^2 = a^2= b^2=1$, $b= \alpha \, a$  and
 $\G{2}= dA+ \G{0} B$ as in (\ref{G2}).
Harmonic functions are defined as follows:
\bea
\label{f}
 f(\vec x) &= &1+\sum_k\frac{N_k\; c^{10}_f}{|\vec x-\vec x_k|^6} \\
 h(z)& = & 1- {n-8\over 2\pi l_s}|z|\,,\,\qquad
\alpha^2=1 \, ,
\label{h}
\eea
where 
\begin{equation}
 c^{10}_f= { 2 \kappa_{10}^2\over  2\pi l_s^2 6 \omega_7}\qquad \omega_7= {2 \pi^{7/2}\over \Gamma(7/2)}\, .
\label{c10}
\end{equation}
The solutions of the equations of motion given above solve the equation of the action $S_{\rm bulk}+ S_{O8D8}$ everywhere but at the positions of the F1-D0 strings at $\vec x = \vec x_k$. The form of the solutions suggest that one can find the source action for F1-D0 strings so that the total action has equation of motion for which (\ref{metr1})-(\ref{c10}) gives a solution everywhere, including the position of the strings at $\vec x = \vec x_k$. We find the appropriate action in the form:
\begin{eqnarray}
S_{F1D0}&=& -\frac{1}{2 \pi \alpha'}
\Big[ \int d^{10}x  \sum_k N_k  \delta^8(\vec x-\vec x_k)\nonumber\\
&&\Big \{ \big(e^{-\phi} \sqrt{-g_{tt}} -b A_t\big)
+\big( \sqrt{-g_{tt}g_{zz}}
- \frac{a}{2!}\epsilon^{\underline \nu \underline \lambda}B_{\underline \nu \underline \lambda}\big)
\frac{e^{-\phi}}{\sqrt{g_{zz}}}\Big\} \Big] \, .
\label{string}
\end{eqnarray}
Here $\underline \nu , \underline \lambda$ take values  $t,z$.
As we see here, the action is not simply related to D0 or F1 solution. The first part somehow reminds a D0 action in a static gauge, however it is integrated over $z$  and not only over $t$ which would correspond to a D0 brane. The second part is almost an F1 string action in a static gauge. However, there is a term  $\frac{e^{-\phi}}{\sqrt{g_{zz}}}$ which for our solution is equal to $h(z)$ and is related to O8-D8 domain wall. This term breaks the $O(1,1)$ symmetry of the fundamental string and is unusual. 

We will present below all equations of motion following from the bulk \& brane action and show that they are solved by (\ref{metr1})-(\ref{c10}). Our total  bulk \& brane action which is a subject for variation over
$
  \left\{   g_{\mu\nu},
  B_{\mu \nu},
  \phi,
  \G{0},
  \A{1}_{\mu},
  \A{9}_{\mu _1\cdots\mu _9 }
  \right\}
 $ fields is:
\bea 
S_{\rm total} &=& S_{\rm bulk}+S_{\rm branes}
= \frac{1}{2\kappa_{10}^2}
\Big[ \int d^{10}x {\cal L}_{bulk} 
\nonumber \\
&-& \tilde{T}_8 \int d^{10}x 
\big(\delta(z)-\delta(z-\pi R)\big)
\big( e^{-\phi}\sqrt{|g_9|}+\alpha 
\frac{1}{9!}\epsilon^{(9)}A^{(9)}\big)
\nonumber \\
&-& \tilde{T}_2 \int d^{10}x
\sum_k N_k \delta^8(\vec x-\vec x_k) 
\Big \{ \big(e^{-\phi} \sqrt{-g_{tt}} -b A_t\big)
\nonumber \\
&+&\big( \sqrt{-g_{tt}g_{zz}}
- \frac{a}{2!}\epsilon^{\underline \nu \underline \lambda}B_{\underline \nu \underline \lambda}\big)
\frac{e^{-\phi}}{\sqrt{g_{zz}}}\Big\} \Big]\label{total} \, .
\eea
The  covariant equations of motion that follow from the bulk action
(\ref{bulk}) are:
\bea
\frac{\delta S}{\delta g^{\mu\nu}}|_{\rm bulk} & = &
   -\frac{\sqrt{-g}}{2\kappa_{10}^2} \Big( e^{-2\phi}
  \Big[ R_{\mu \nu}-\frac{1}{2}g_{\mu \nu}R
   + 2 g_{\mu \nu}{\nabla }^2 \phi - 2 \nabla_{\mu} \nabla_{\nu} \phi
   \nonumber \\
   &\, &- 2g_{\mu \nu}\partial^\sigma \phi \partial_\sigma \phi
   - \frac{1}{24}g_{\mu \nu}H_{\lambda \rho \sigma}
   H^{\lambda \rho \sigma}
   + \frac{1}{4}H_{\lbrace \mu \lambda \sigma}
    H_{\nu \rbrace}^{\lambda \sigma}\Big] \nonumber \\
   &\, &
   -\frac{1}{4}g_{\mu \nu}(\G{0})^2
   -\frac{1}{8}g_{\mu \nu}(\G{2})_{\lambda \sigma}
    (\G{2})^{\lambda \sigma}
    +\frac{1}{2}(\G{2})_{\lbrace \mu \lambda }
    (\G{2})_{\nu \rbrace}^{\lambda} \Big)  \label{Einsteq} \\
   \frac{\delta S}{\delta \phi}|_{\rm bulk} & = &
    -\frac{\sqrt{-g}}{\kappa_{10}^2}  e^{-2\phi}
   \Big(
   4\nabla^2 \phi - 4\partial_\mu \phi \partial^\mu \phi
  -R -\frac{1}{12}H_{\lambda \rho \sigma}
   H^{\lambda \rho \sigma}  \Big) \label{phieq} \\
  \frac{\delta S}{\delta B^{\mu\nu}}|_{\rm bulk} & = &
   -\frac{\sqrt{-g}}{2\kappa_{10}^2}
   \Big( - D^\lambda \big( e^{-2\phi}\frac{1}{2}
   H_{\mu \nu \lambda}\big) +
  \frac{1}{2} \G{0}(\G{2})_{\mu \nu} \Big )
    \label{beq} \\
  \frac{\delta S}{\delta A^{(1)}_\nu}|_{\rm bulk} & = &
  -\frac{\sqrt{-g}}{2\kappa_{10}^2}
  \Big( - D_\mu(\G{2})^{\mu \nu} \Big) \label{a1eq} \\
\frac{\delta S}{\delta A^{(9)}_{\mu_1 \cdots \mu_{9}}}|_{\rm bulk} & = &
  -\frac{1}{2\kappa_{10}^2}\Big( - \frac{1}{9!}
 \epsilon^{(10) \mu_1 \cdots \mu_{9}\nu}
\partial_{\nu} \G{0} \Big) \label{a9eq} \\
\frac{\delta S}{\delta \G{0}}|_{\rm bulk} & = &
-\frac{\sqrt{-g}}{2\kappa_{10}^2} \Big(
\G{0}+\frac{1}{2}(\G{2})_{\mu \nu}B^{\mu \nu}
-\frac{1}{9!\sqrt{-g}}
 \epsilon^{(10) \mu_1 \cdots \mu_{10}}
\partial_{\mu_1}A^{(9)}_{\mu_2 \cdots \mu_{10}} \Big)
\label{goeq}
 \eea
It is easy to see that the last equation
becomes a standard duality equation  when
$(\G{2})_{\mu \nu}B^{\mu \nu}
=0$, which is indeed a property of our solution 
(\ref{metr1})-(\ref{c10}).
When our ansatz (\ref{metr1})-(\ref{g21}) is substituted in bulk equations of motion, we find a wonderful simplification
 \bea
\frac{\delta S}{\delta g^{tt}}|_{\rm bulk}
&= & \frac{g_{tt}}{2\kappa_{10}^2} \Big( 
h f^{-1}{\bf \Delta} f 
+\frac{1}{2}h^{-1} f\partial_z 
\partial_z h
\Big )\, , 
\label{ggtt} \\
\frac{\delta S}{\delta g^{zz}}|_{\rm bulk}
& = &  \frac{g_{zz}}{4\kappa_{10}^2} \Big( 
h f^{-1}{\bf \Delta} f 
\Big)\, , 
\label{ggzz} \\
\frac{\delta S}{\delta g^{\rm ii}}|_{\rm bulk}
 & = & \frac{g_{\rm ii}}{4\kappa_{10}^2} \Big( 
h^{-1} 
f\partial_z 
\partial_z h 
\Big)\, , \label{ggii} \\
\frac{\delta S}{\delta \phi}|_{\rm bulk} & =  &
  \frac{1}{2\kappa_{10}^2}\big(
  h f^{-1}{\bf \Delta} f  
+ h^{-1} f \partial_z 
  \partial_z h \big)\, , 
 \label{gphivar} \\
\frac{\delta S}{\delta B_{zt}}|_{\rm bulk}  & = & 
  \frac{a}{4\kappa_{10}^2}h {\bf \Delta} f  \, , 
\label{gbeq} \\
\frac{\delta S}{\delta A^1_t}|_{\rm bulk} & = & 
 \frac{1}{2\kappa_{10}^2}\frac{a}{\alpha}
 {\bf \Delta} f \, ,\\
 \label{ga1var} 
 \frac{\delta S}{\delta A^{(9)}_{\mu_1 \cdots \mu_{9}}}|_{\rm bulk}&=& -\frac{\alpha }{2\kappa_{10}^2} 
\frac{\epsilon^{(9)}}{9!}\Big\{
 \partial_{z}\partial_{z}h(z)
 \Big\}\, ,
 \\
\frac{\delta S}{\delta \G{0}}|_{\rm bulk} & = &
-\frac{\sqrt{-g}}{2\kappa_{10}^2} \Big(
\G{0} + \star  \G{10} \Big)\, .
\eea
In the presence of sources introduced in eq. (\ref{total}) the complete equations of motion have additional terms:
 \bea
\frac{\delta S}{\delta g^{tt}}|_{\rm bulk\& brane}
&= & \frac{g_{tt}}{2\kappa_{10}^2} \Big \{ 
h f^{-1}{\bf \Delta} f  + 
\tilde{T}_2h f^{-1}
\sum_k N_k \delta^8(\vec x-\vec x_k)
  \Big\} \nonumber \\
& + &  \frac{g_{tt}}{4\kappa_{10}^2} 
\Big \{h^{-1} f\partial_z 
\partial_z h + 
\tilde{T}_8 h^{-1} f
\big(\delta(z)-\delta(z-\tilde{z})\big)
 \Big\} =0
\, , 
\label{bggtt} \\
\frac{\delta S}{\delta g^{zz}}|_{\rm bulk\& brane}
& = &  \frac{g_{zz}}{4\kappa_{10}^2} \Big\{ 
h f^{-1}{\bf \Delta} f 
+\tilde{T}_2 h f^{-1}
\sum_k N_k \delta^8(\vec x-\vec x_k) 
\Big\}=0
\, , 
\label{bggzz} \\
\frac{\delta S}{\delta g^{\rm ii}}|_{\rm bulk\& brane}
& = & \frac{g_{\rm ii}}{4\kappa_{10}^2} \Big\{
h^{-1} f\partial_z 
\partial_z h 
+\tilde{T}_8 h^{-1} f
\big(\delta(z)-\delta(z-\tilde{z})\big)
 \Big\} =0
\, , \label{bggii} \\
\frac{\delta S}{\delta \phi}|_{\rm bulk\& brane} 
& = & \frac{1}{2\kappa_{10}^2}\Big\{
  h f^{-1}{\bf \Delta} f  +
\tilde{T}_2 h f^{-1}
\sum_k N_k \delta^8(\vec x-\vec x_k)
\Big\} \nonumber \\
&+& \frac{1}{2\kappa_{10}^2}\Big\{
 h^{-1} f \partial_z 
  \partial_z h
+ \tilde{T}_8 h^{-1} f
\big(\delta(z)-\delta(z-\tilde{z})\big)=0
\, , 
 \label{bgphivar} \\
\frac{\delta S}{\delta B_{zt}}|_{\rm bulk\& brane}   
& = & 
  \frac{a}{4\kappa_{10}^2}\Big\{h {\bf \Delta} f 
+\tilde{T}_2 h
\sum_k N_k \delta^8(\vec x-\vec x_k)
 \Big\}=0 
\, , 
\label{bgbeq} \\
\frac{\delta S}{\delta A^1_t}|_{\rm bulk\& brane} & = & 
 \frac{1}{2\kappa_{10}^2}\frac{a}{\alpha}
 \Big\{{\bf \Delta} f +  
\tilde{T}_2 
\sum_k N_k \delta^8(\vec x-\vec x_k) \Big\}=0 \, ,
 \label{bga1var} \\
\frac{\delta S}{\delta A^{(9)}_{\mu_1 \cdots \mu_{9}}}|_{\rm bulk\& brane} 
& = &
  -\frac{\alpha }{2\kappa_{10}^2} 
\frac{\epsilon^{(9)}}{9!}\Big\{
 \partial_{z}\partial_{z}h +
\tilde{T}_8 
\big(\delta(z)-\delta(z-\tilde{z})\big)
 \Big\}=0 \, ,
\label{ba9eq} \\
\frac{\delta S}{\delta \G{0}}|_{\rm bulk\& brane} & = &
-\frac{\sqrt{-g}}{2\kappa_{10}^2} \Big(
\G{0} + \star  \G{10} \Big)=0 \, .
\eea
All equations of motion of bulk \& brane action are miraculously satisfied under condition that the equations for the 
 harmonic functions $f, h$ include the domain wall and multi-string source terms:
\bea
{\bf \Delta} f  +\tilde{T}_2 
\sum_k N_k \delta^8(\vec x-\vec x_k) = 0 
\label{fequ} \\ 
\partial_{z}\partial_{z}h +
\tilde{T}_8 
\big(\delta(z)-
\delta(z-\pi R)\big)=0 \, .
\label{hequ} 
\eea
These equations are satisfied by our harmonic functions defined in  (\ref{f})-(\ref{c10}).

\

In conclusion, we have found a 1/4 BPS solution of IIA (massive) theory with domain walls at the fixed points of the orientifold and multiple strings stretched between domain walls. The configuration has some electric field, $A_t(z, \vec x)$ reminiscent of the D0-brane and some 2-form $B_{zt}(\vec x)$ reminiscent of the F1 multi-string solution. There is a piecewise constant 0-form, dual to a 10-form, and both change the sign across the wall. We leave it to future investigations to find a better interpretation of this configuration and to understand the possibilities to use it. 

\

Acknowledgments. We had useful discussions with  O. Bergman, E. Halyo,
S. Hellerman, I. Klebanov, A. Linde, J. McGreevy, A. Peet and J. Polchinski. We are grateful to
Y. Zunger for his GRONK program\footnote{http://itp.stanford.edu/~zunger/} which we used for calculations.
The work is supported by NSF grant PHY-9870115. The work of M.S. was partly supported by  the Department of Energy under a contract DE-AC03-76SF00515. S.P. acknowledges the support from Stanford Graduate Fellowship
Foundation.

\section*{ Appendix: 1/4 of Unbroken Supersymmetry }

Our solution has an $SO(8)$ symmetry. For simplicity we will  switch to polar coordinates and consider one string solution at $(\vec x)^2= r^2=0$. We will find that $\delta \lambda=0$ and $\delta \psi_t= \delta \psi_z= \delta \psi_r=0$. Due to $SO(8)$ symmetry the variation of the gravitino $\delta \psi_i$ can be written as follows $\delta \psi_i= \delta_{ij} x^j \Psi $. The relation to $\delta \psi_r$ is 
\begin{equation}
 \delta \psi_r= \delta \psi_i {x^i\over r}= {(x^i)^2\over r} \Psi= \Psi
\label{polar}
\end{equation}
Thus if we establish that $ \delta \psi_r=0$ it will follow that $\delta \psi_i= \delta_{ij} x^j \Psi =0 $.

We will substitute our solution into the supersymmetry
transformations for $\lambda $:
\bea
\delta_{{\epsilon}}{\lambda} & =
    & \Big( \! \! \not \! \partial \phi
    - {\textstyle\frac{1}{12}}
    {\Gamma}_{11} \not\!\! {H} \Big)
      {\epsilon}
    + {\textstyle\frac{1}{4}} e^{{\phi}}
     \Big(5 {G}^{(0)}
    + {\textstyle\frac{3}{2}}
    (2 \, \partial_{[\nu}\A{1}_{\rho]}
    {\Gamma}^{\nu \rho}
    +\G{0}\not\!B){\Gamma}_{11}
     \Big){\epsilon}  \nonumber
\eea
and use for our ansatz that
\bea
 \not \! \partial \phi
  &=& {\Gamma}^{\mu}\partial_{\mu}\phi
  =  \frac{1}{4}h^{\frac{1}{4}}
  f^{-\frac{5}{4}}\partial_r f
  {\Gamma}^{ r }
  -\frac{5}{4}h^{-\frac{5}{4}}
  f^{\frac{1}{4}}\partial_z h
  {\Gamma}^{\underline{z}} \nonumber \\
  - {\textstyle\frac{1}{12}} {\Gamma}_{11} \not\!\! {H}
  &=& -\frac{\alpha a}{2}h^{\frac{1}{4}}
  f^{-\frac{5}{4}}\partial_r f
  {\Gamma}^{rzt}{\Gamma}^{11} \nonumber \\
  {\textstyle\frac{5}{4}} e^{{\phi}}{G}^{(0)}
  &=& -{\textstyle\frac{5}{4 \alpha }}
  h^{- \frac{5}{4}}
  f^{\frac{1}{4}}\partial_z h
  \nonumber \\
  {\textstyle\frac{3}{4}}e^{{\phi}}{G}^{(2)}_{rt}
  {\Gamma}^{rt}{\Gamma}^{11}
  &=&-{\textstyle\frac{3a}{4}}h^{\frac{1}{4}}
  f^{-\frac{5}{4}}\partial_r f
  {\Gamma}^{rt}{\Gamma}^{11}. \nonumber
  \eea
Collecting all terms we have:
\bea
\delta_{{\epsilon}}{\lambda}
   = \frac{1}{4}h^{\frac{1}{4}}
  f^{-\frac{5}{4}}\partial_r f
  {\Gamma}^{ r }\big(1-2a \alpha {\Gamma}^{zt}
   {\Gamma}^{11}
   -3a{\Gamma}^{t}{\Gamma}^{11}\big){\epsilon}
  - {\textstyle \frac{5}{4 \alpha}} h^{-\frac{5}{4}}
  f^{\frac{1}{4}}
  \partial_z h\big(1
  + \alpha {\Gamma}^{z}\big){\epsilon} =0\,.
  \label{susyla}
\eea
We impose the following projectors:
\bea
  \alpha  {\Gamma}^{z}{\epsilon}
  &= &-{\epsilon}\nonumber \\
  \alpha {\Gamma}^{zt}{\Gamma}^{11}{\epsilon}
  &=& \beta {\epsilon}\,, \quad \beta^2=1
  \label{project}\\
  {\Gamma}^{t}{\Gamma}^{11}{\epsilon}
  &=& \gamma {\epsilon}\,, \quad \gamma^2=1
  \nonumber
 \eea
 The first one is for D8, the second one is related to F1 and the third one (a product of the first two) is related to D0. Thus we have 1/4 of supersymmetry unbroken, since only two projectors are independent.

The dilatino transformation (\ref{susyla}) vanishes under conditions that $a(2\alpha \beta + 3\gamma) =1$.
Therefore there are two possibilities:
\bea
  {\bf (1)} \quad  \alpha \beta = 1\,
  ( \alpha =\pm 1, \beta = \pm 1)\,,
  \quad \gamma = -1\,\, &\Rightarrow &\,
  a= -1 \nonumber \\
  {\bf (2)} \quad \alpha \beta = -1\,
  ( \alpha =\mp 1, \beta = \pm 1)\, ,
   \quad \gamma = 1\,\,
    &\Rightarrow  &\, a= 1 \nonumber
\eea
and it is easy to check the compatibility of
these projectors.

In our next step we have to consider the supersymmetry
 transformations for  ${\psi}$
\bea
\delta_{{\epsilon}} {\psi}_{{\mu}} & =
  & \Big( \partial_{{\mu}} +{\textstyle\frac{1}{4}}
    \not\!{\omega}_{{\mu}}
    +{\textstyle\frac{1}{8}}\Gamma_{11}\not\!\! {H}_{\mu}
    \Big) {\epsilon}
    +{\textstyle\frac{1}{8}} e^{{\phi}} \Big({G}^{(0)}
    {\Gamma}_{{\mu}}
    +{\textstyle\frac{1}{2}}
    (2 \, \partial_{[\nu}\A{1}_{\rho]}
    {\Gamma}^{\nu \rho}
     +\G{0}\not\!B){\Gamma}_{{\mu}} {\Gamma}_{11}
        \Big){\epsilon} \nonumber
\eea
and we assume  that
${\epsilon} =(-g_{tt})^{\frac{1}{4}} {\epsilon}_0.$ (In polar coordinates 
${\epsilon}_0$ depends on angles). The set of useful expressions is:
\bea
   e^t_{\underline t}=h^{-\frac{1}{4}}
   f^{-\frac{3}{4}}\,, \, &\quad &\,
   \not\!{\omega}_{t} =  -\frac{3}{2}
   f^{ -2}\partial_r f {\Gamma}^{rt}
     -\frac{1}{2} h^{-\frac{3}{2}}
   f^{ -\frac{1}{2}}\partial_z
   h {\Gamma}^{zt}\nonumber \\
   e^z_{\underline z}=h^{\frac{1}{4}}
   f^{-\frac{1}{4}}\,, \, &\quad &\,
   \not\!{\omega}_{z} = \frac{1}{2}
    h^{\frac{1}{2}}
   f^{ -\frac{3}{2}}\partial_r
   f {\Gamma}^{rz}
   \nonumber \\
   e^r_{\underline r}=h^{-\frac{1}{4}}
   f^{\frac{1}{4}}\,, \, &\quad &\,
    \not\!{\omega}_{r} =
   \frac{1}{2} h^{-\frac{3}{2}}
   f^{ \frac{1}{2}}\partial_z
   h {\Gamma}^{zr} \nonumber
\eea

 Collecting all terms we  get:
\bea
  \delta_{{\epsilon}} {\psi}_{t} & = &
  - \frac{1}{8}
    f^{-2}\partial_r
    f {\Gamma}^{r}\Big(3{\Gamma}^{t}
  + 2\alpha a {\Gamma}^{z}{\Gamma}^{11}
  +  a {\Gamma}^{11}\Big){\epsilon}
  + \frac{1}{8\alpha }
   h^{-\frac{3}{2}}
   f^{ -\frac{1}{2}}\partial_z
   h {\Gamma}^{t}\Big(1
  + \alpha {\Gamma}^{z}\Big){\epsilon} \nonumber \\
 \delta_{{\epsilon}} {\psi}_{z} & = &
  \frac{1}{8} h^{\frac{1}{2}}
   f^{ -\frac{3}{2}}\partial_r
   f {\Gamma}^{r}\Big({\Gamma}^{z}
  + 2\alpha a {\Gamma}^{t}{\Gamma}^{11}
  +  a{\Gamma}^{zt}{\Gamma}^{11}\Big){\epsilon}
  - \frac{1}{8\alpha }
   h^{-1}\partial_z h \Big(\alpha
  +  {\Gamma}^{z}\Big){\epsilon} \nonumber \\
  \delta_{{\epsilon}} {\psi}_{r} & = &
  - \frac{1}{8 }  f^{-1}\partial_r f
  \Big(3
  + 2\alpha a {\Gamma}^{zt}{\Gamma}^{11}
  -  a{\Gamma}^{t} {\Gamma}^{11}\Big){\epsilon}
  - \frac{1}{8\alpha }
   h^{-\frac{3}{2}}
   f^{\frac{1}{2}}\partial_z
   h {\Gamma}^{r}\Big(1
  + \alpha {\Gamma}^{z}\Big){\epsilon} \nonumber
\eea
It is easy to show that if  the spinors $\epsilon$ satisfy the projector
conditions
 (\ref{project})  
$$\delta {\psi}_{t}=\delta {\psi}_{z}= \delta {\psi}_{r}=0 .$$ As we explained above, it also means that 
$$\delta {\psi}_{i}=0 \,  \qquad i=1, \dots , 8.$$
We have shown that our  solution (\ref{metr1})-(\ref{g21})
 satisfies the  condition of 1/4 of unbroken supersymmetry.

\end{document}